\documentclass[11pt,a4paper]{article}
\pdfoutput=1
\usepackage{jheppub}

\usepackage{color}
\usepackage{amsmath}
\usepackage{pifont}   
\usepackage{verbatim}       
\usepackage{acronym} 
    
\usepackage{amsfonts}
\usepackage{amssymb}
\usepackage{mathrsfs} 
\usepackage{graphicx}
\usepackage{multirow} 
\usepackage{slashed} 
\usepackage{array}

\usepackage{graphicx}
\usepackage{epsfig}
\usepackage{lscape}
\usepackage{rotating}
\usepackage{epsf}
\usepackage{bm}
\usepackage{booktabs}
\usepackage{array}
\usepackage{caption}
\usepackage{subcaption}
\usepackage[utf8]{inputenc}
\usepackage{float}

\usepackage{multirow}
\usepackage{array,booktabs,colortbl,multirow}
\usepackage{colortbl,xcolor,colordvi,color}
\usepackage{rotating}
\allowdisplaybreaks

\newcommand{\cmrule}{\midrule[0.25mm]}

\title{Mirror neutrons as dark matter in the Mirror Twin Two Higgs Doublet Model}

\author{Hugues Beauchesne}

\affiliation{Department of Physics, Ben-Gurion University, \\Beer-Sheva 8410501, Israel}

\emailAdd{beauches@post.bgu.ac.il}

\abstract{In addition to being a solution to the little hierarchy problem, the Mirror Twin Higgs provides a natural setting for Asymmetric Dark Matter. In its incarnation with only one Higgs doublet and its mirror copy, dark matter would however almost certainly consist mostly of mirror atoms, which is severely ruled out by constraints on dark matter self-interactions. By adding a second Higgs doublet and its mirror, the vevs of the different Higgses can be arranged such that dark matter consists mostly of mirror neutrons, which is cosmologically viable. In this paper, it is shown that current constraints from colliders, flavour and cosmology can accommodate such a vev structure with little increase in the necessary tuning.}

\begin{document}

\maketitle

\section{Introduction}\label{Sec:Introduction}
The Twin Higgs \cite{Chacko:2005pe, Barbieri:2005ri} is an attempt to solve the little hierarchy problem by introducing partners that are neutral under the Standard Model (SM) gauge groups, being an example of so-called Neutral Naturalness. Its conceptually simplest form is the Mirror Twin Higgs. In this model, a mirror copy of the Standard Model is introduced, including a copy of every SM particle and the equivalent interactions between them. The main difference is that the new particles are charged under new mirror gauge groups that reflect those of the Standard Model. In other words, the Twin partner of a SM particle is charged under the mirror $SU(3)\times SU(2) \times U(1)$ in the same way that said SM particle is charged under the $SU(3)\times SU(2) \times U(1)$ of the Standard Model. The Higgs doublet and its mirror partner can then be combined to write a potential that respects an approximate $SU(4)$ global symmetry, which is assumed to be broken spontaneously to $SU(3)$. This results in seven (pseudo-)Goldstone bosons, three of which are eaten by SM gauge bosons and three by the corresponding mirror gauge bosons. The remaining one serves as a candidate for the observed Higgs particle. Central to the Twin Higgs model is the introduction of a $\mathbb{Z}_2$ interchange symmetry between the SM and mirror sectors. This symmetry insures that the leading corrections to the potential, the ones proportional to the cutoff of the theory square, respect the $SU(4)$ global symmetry. The Higgs being a pseudo-Goldstone boson of this symmetry, it does not gain a mass from these leading corrections by themselves and its mass is effectively protected at one-loop by the mirror partners. This interchange symmetry imposes an equality between the gauge and Yukawa couplings of the two sectors, but must be broken to be experimentally viable. Without any hard breaking, respecting current constraints and avoiding too much tuning results in the mirror partners only being a factor of a few heavier than their SM counterparts.

The cosmology of the Mirror Twin Higgs is both promising and challenging. On one hand, it naturally accommodates Asymmetric Dark Matter \cite{Petraki:2013wwa, Kaplan:2009ag, Zurek:2013wia}. Since the observable Universe consists overwhelmingly of matter and not antimatter, there must have been some sort of baryogenesis mechanism at work during early times. Since the mirror sector is related to the SM sector by an interchange symmetry, it is not unreasonable to think that it would exhibit a similar asymmetry. If the mirror baryons were a bit heavier or their numbers were slightly larger, they would have an abundance that matches naturally with the observed abundance of dark matter. This provides a very elegant explanation for dark matter, which has been explored in Refs.~\cite{Farina:2015uea, Farina:2016ndq, Garcia:2015toa, Feng:2020urb}.  On the other hand, the cosmology of the Mirror Twin Higgs faces two major challenges. First, the presence of a mirror photon and mirror neutrinos leads to too many effective relativistic degrees of freedom, which is severely constrained by both the Cosmic Microwave Background (CMB) and Big Bang Nucleosynthesis (BBN) \cite{Fields:2019pfx, Aghanim:2018eyx}. This problem is not the focus of this paper, but we do mention that several solutions were already proposed and involve cooling down the mirror sector before the onset of BBN \cite{Chacko:2016hvu, Craig:2016lyx, Csaki:2017spo, Harigaya:2019shz, Koren:2019iuv}. Second, the $\mathbb{Z}_2$ symmetry only being broken softly would almost unavoidably lead to dark matter having a similar composition to normal matter, i.e. mostly mirror hydrogen and mirror helium. The problem with this is that the dark matter self-interaction would be similar to that between regular atoms. If mirror atoms were to represent most of the dark matter, their self-interaction cross sections would be ruled out by several orders of magnitude or would require making the mirror atoms far more compact or heavy than can naturally be accommodated in the Mirror Twin Higgs \cite{Kaplan:2009de, CyrRacine:2012fz, Cline:2013pca}. See also for example Refs.~\cite{Hodges:1993yb, Berezhiani:2000gw, Foot:2004pa, An:2009vq, Roux:2020wkp, khlopov1, khlopov2, khlopov3, khlopov4, Foot:2014uba, Foot:2016wvj} for dark matter in Mirror World.

This begs the question of whether it is possible for dark matter to take an acceptable form in the Mirror Twin Higgs realization of Asymmetric Dark Matter. A simple solution would be if dark matter consisted mainly of mirror neutrons. These would have acceptable self-interactions and might even solve some of the small scale structure problems \cite{Spergel:1999mh}. This possibility was considered in Refs.~\cite{Farina:2015uea, Barbieri:2016zxn, Hall:2019rld}, albeit with hard $\mathbb{Z}_2$ breaking.

A simple way to have the dark matter consist mostly of mirror neutrons without introducing any hard breaking of the $\mathbb{Z}_2$ symmetry is instead to have a Two Higgs Doublet Model and a copy of it, or a so-called Mirror Twin Two Higgs Doublet Model (MT2HDM). In any case, the $\mathbb{Z}_2$ symmetry must be broken to pass the Higgs signal strengths constraints. Even some soft or spontaneous breaking will in general result in a $\tan\beta$ in the mirror sector that differs from its SM value. As such, the Yukawa couplings might be the same in the two sectors, but the ratio of quark masses can be considerably different in the SM and mirror sectors. By making the ratio of the mirror up and down quarks sufficiently large, the mirror proton can be made heavier than the mirror neutron. This results in at least a reduced abundance of dark protons, but also the possibilities of free mirror protons being unstable or even mirror protons being unstable in all the bound states with sizable abundances. These possibilities can in principle result in the dark matter being mainly mirror neutrons, with maybe an experimentally acceptable amount of mirror atoms or dark hadrons made of multiple mirror neutrons. More complex vev structures in the MT2HDM have already been studied in Refs.~\cite{Beauchesne:2015lva, Harnik:2016koz, Yu:2016bku, Yu:2016swa, Katz:2016wtw}. Note also that this possibility was considered in the context of Mirror World \cite{Addazi:2015cua} and was mentioned in passing for the Twin Higgs in Ref.~\cite{Chacko:2018vss}.

Though a simple and attractive idea, it is not at first trivial that having the dark matter consist mainly of mirror neutrons in the MT2HDM will not introduce new problems or if it is even possible. It introduces new particles for which there exists both collider and flavour constraints. The Higgs signal strengths constraints must still be satisfied. The cosmology must be viable. Also crucially, the Twin Higgs is supposed to be a solution to the little hierarchy problem. Making the mirror neutron into a valid dark matter candidate should not introduce an unreasonable amount of tuning.

With this context in mind, the goal of this paper is to determine whether mirror neutrons can be made into valid dark matter candidates in the MT2HDM and, if so, what amount of tuning is necessary. To do so, this paper will look at current constraints from colliders, flavour and cosmology.

The end result will be that the MT2HDM can indeed accommodate mirror neutron dark matter. The ability to do so is however highly dependent on the masses of the up and down quarks and it was not trivial beforehand that the currently measured values would be acceptable. Small fluctuations of even one sigma around the central values of these measurements can make obtaining viable dark matter range from being almost trivial to only possible in a very narrow region of parameter space. It will also be demonstrated that more stringent constraints on the Higgs signal strengths and the effective number of relativistic degrees of freedom could drastically modify the viability of this scenario. Finally, it will be shown that there would often be a non-negligible amount of mirror atoms in the dark matter, for which there exists interesting signatures~\cite{Chacko:2018vss, Curtin:2019lhm, Curtin:2019ngc}.

The article is organized as follows. First, the full potential is introduced and the general idea as to how to reduce the amount of dark atoms is explained. Next, the collider and flavour constraints are discussed and constraints on the different vevs are presented. Then, the cosmological evolution and astrophysical constraints are discussed. Finally, a series of plots assembling all the results are presented.

\section{Potential, notation and basic concepts}\label{Sec:Potential}
We begin by presenting the potential that will be used throughout the paper and explain the general idea as to how to obtain mirror neutron dark matter. Some useful notation is also introduced.

As is common practice, the Standard Model particles are labelled with an $A$ and their mirror partners with a $B$. Two Higgs doublets of weak hypercharge $+1/2$ are introduced in the SM sector, $H_1^A$ and $H_2^A$, and their mirror partners are respectively $H_1^B$ and $H_2^B$. A SM Higgs doublet and its mirror partner can then be combined to form a fundamental representation of the global $SU(4)$ as $H_i^T = ((H_i^A)^T, (H_i^B)^T)$. The general renormalizable potential that we will consider is
\begin{equation}\label{eq:GeneralPotential1}
  \begin{aligned}
    V = &-\mu_1^2|H_1|^2  -\mu_2^2|H_2|^2 + \lambda_1|H_1|^4 +\lambda_2|H_2|^4 
         +  \alpha_1|H_1^A|^2|H_1^B|^2           + \alpha_2|H_2^A|^2|H_2^B|^2       \\
        &+  \Delta m_1^2 |H_1^A|^2               +  \Delta m_2^2 |H_2^A|^2  - \left(B_\mu H_1^\dagger H_2 + \text{h.c.}\right) 
         -  \left(\Delta B_\mu (H_1^A)^\dagger H_2^A + \text{h.c.}\right)\\
        &+ \lambda_3|H_1|^2|H_2|^2 + \lambda_4 H_1^\dagger H_2 H_2^\dagger H_1 + \left[\lambda_5(H_1^\dagger H_2)^2 + \text{h.c.}\right]    \\
        &+ \beta_3\left[|H_1^A|^2|H_2^A|^2 + |H_1^B|^2|H_2^B|^2\right]                                                                      \\
        &+ \beta_4\left[((H_1^A)^\dagger H_2^A)((H_2^A)^\dagger H_1^A) + ((H_1^B)^\dagger H_2^B)((H_2^B)^\dagger H_1^B)\right]              \\
        &+\left[ \beta_5\left(((H_1^A)^\dagger H_2^A)^2 + ((H_1^B)^\dagger H_2^B)^2\right) + \text{h.c.}\right].\\        
  \end{aligned}
\end{equation}
Several comments are in order. The conventions are based on Ref.~\cite{Beauchesne:2015lva} with minor modifications. The potential $V$ respects the $\mathbb{Z}_2$ interchange symmetry up to soft-breaking. In addition, it respects an additional softly-broken $\mathbb{Z}_2$ symmetry under which $H_1$ is even and $H_2$ odd. This prevents tree-level flavour changing neutral currents. All parameters are assumed real such that the potential preserves CP symmetry. Other terms can be written, but can be shown to be linear combinations of those above.

The Higgs multiplets will in general acquire expectation values. Assuming no spontaneous charge or CP-breaking in any sector, these can be written in the form
\begin{equation}\label{eq:vevsDef}
  \langle H_1 \rangle = \frac{v_1}{\sqrt{2}}\begin{pmatrix} 0 \\ \sin\theta_1 \\ 0 \\ \cos\theta_1 \end{pmatrix} = \frac{1}{\sqrt{2}} \begin{pmatrix} 0 \\ v_1^A \\ 0 \\ v_1^B \end{pmatrix}, \qquad
  \langle H_2 \rangle = \frac{v_2}{\sqrt{2}}\begin{pmatrix} 0 \\ \sin\theta_2 \\ 0 \\ \cos\theta_2 \end{pmatrix} = \frac{1}{\sqrt{2}} \begin{pmatrix} 0 \\ v_2^A \\ 0 \\ v_2^B \end{pmatrix}.
\end{equation}
The up-type quarks are assumed to obtain a mass from $H_2$ and the down-type quarks from $H_1$. Their masses respect at leading order
\begin{equation}\label{eq:FermionMassesd}
  \frac{m_{q^B}}{m_{q^A}} = \frac{v_1^B}{v_1^A} = \frac{1}{\tan\theta_1} \text{\hspace{0.5cm}for } q \in \{d, s, b\},
\end{equation}
\begin{equation}\label{eq:FermionMassesu}
  \frac{m_{q^B}}{m_{q^A}} = \frac{v_2^B}{v_2^A} = \frac{1}{\tan\theta_2} \text{\hspace{0.5cm}for } q \in \{u, c, t\}.
\end{equation}
The notation of $m_p$ for the mass of a particle $p$ will be used throughout the paper. For the origin of lepton masses, both Type II and Type Y (flipped) will be considered. The equivalent of Eq.~\eqref{eq:FermionMassesd} applies for the former case and the equivalent of Eq.~\eqref{eq:FermionMassesu} for the latter. The masses of the mirror electroweak bosons respect at leading order
\begin{equation}\label{eq:VMmasses}
  \frac{m_{W^B}}{m_{W^A}} = \frac{m_{Z^B}}{m_{Z^A}} = \frac{v^B}{v^A},
\end{equation}
where $v^A = \sqrt{(v_1^A)^2 + (v_2^A)^2}$ and $v^B = \sqrt{(v_1^B)^2 + (v_2^B)^2}$.

One of the obvious consequences of both Two Higgs Doublet Models and the Twin Higgs is that the different Higgs mass eigenstates are in general a linear combination of different gauge eigenstates. Label the CP-even neutral part of $H_i^M$ as $h_i^M$, the CP-odd part as $a_i^M$ and the charged part as ${H^+_i}^M$. Mass eigenstates and gauge eigenstates are then related via
\begin{equation}\label{eq:Mixing}
  \begin{pmatrix} h_1 \\ h_2 \\ h_3 \\ h_4 \end{pmatrix} = R^e\begin{pmatrix} h_1^A \\ h_2^A \\ h_1^B \\ h_2^B \end{pmatrix},\qquad
  \begin{pmatrix} a_1 \\ a_2 \\ a_3 \\ a_4 \end{pmatrix} = R^o\begin{pmatrix} a_1^A \\ a_2^A \\ a_1^B \\ a_2^B \end{pmatrix},\qquad
  \begin{pmatrix} H^+_1 \\ H^+_2 \\ H^+_3 \\ H^+_4 \end{pmatrix} = R^+\begin{pmatrix} {H^+_1}^A \\ {H^+_2}^A \\ {H^+_1}^B \\ {H^+_2}^B \end{pmatrix},
\end{equation}
where $R^e$, $R^o$ and $R^+$ are orthogonal matrices and the $h_i$'s, $a_i$'s and $H^+_i$'s are mass eigenstates ordered from lightest to heaviest. Obviously, $a_i$ and $H^+_i$ are Goldstone bosons for $i$ corresponding to 1 or 2 and there is no mixing between charged states of the two sectors.

Finally, mirror BBN depends on the potential  essentially only insofar as it determines the vevs of the four Higgs doublets. These in turn determine the mirror Fermi constant, masses of the mirror electron, up quark and down quark and, via modifying the running, the mirror QCD scale. Since $v^A$ is known from the mass of the Z boson, only three vevs are independent. These can be expressed in terms of the convenient basis
\begin{equation}\label{eq:Basis}
  \frac{v^B}{v^A}, \qquad \frac{\tan\beta^B}{\tan\beta^A}, \qquad \tan\beta^A,
\end{equation}
where $\tan\beta^A = v_2^A/v_1^A$ and $\tan\beta^B = v_2^B/v_1^B$. The first quantity $v^B/v^A$ determines the mirror Fermi constant and thus affects crucially the mirror temperature at which conversion of mirror protons to mirror neutrons freezes-out. Obviously, a small $v^B/v^A$ can help minimize the amount of mirror atoms, but there is in practice a lower limit on this ratio from Higgs signal strengths measurements. The second quantity $\tan\beta^B/\tan\beta^A$ is crucial as the ratio of the masses of the mirror up and down quark can be written in the convenient form
\begin{equation}\label{eq:mirrorQuarkMassRatio}
  \frac{m_{u^B}}{m_{d^B}} = \frac{\tan\beta^B}{\tan\beta^A}\frac{m_{u^A}}{m_{d^A}} = \frac{\tan\theta_1}{\tan\theta_2}\frac{m_{u^A}}{m_{d^A}}.
\end{equation}
This means that a $\tan\beta^B/\tan\beta^A$ of at least $\sim$2 or preferably larger is required. In simple terms, to obtain a large $\tan\beta^B/\tan\beta^A$, it suffices to make $\langle H_2 \rangle$ more aligned with the $B$ sector than $\langle H_1 \rangle$. A large $m_{u^B}/m_{d^B}$ can then be obtained either by aligning $\langle H_2 \rangle$ close to the $B$ sector ($\tan\theta_2\to 0$) or by aligning $\langle H_1 \rangle$ close to the $A$ sector ($\tan\theta_1\to \infty$). The first option will need to be pursued in any case to pass the Higgs signal strengths requirements. However, simply aligning $\langle H_2 \rangle$ close to the $B$ sector will sometimes prove to be insufficient and it will therefore be necessary to also align $\langle H_1 \rangle$ close to the $A$ sector, which in these cases will introduce additional tuning. Finally, $\tan\beta^A$ is important because Eqs.~\eqref{eq:FermionMassesd} and \eqref{eq:FermionMassesu} can be rewritten in terms of the parameters of Eq.~\eqref{eq:Basis} as
\begin{equation}\label{eq:mdtanbeta}
    \frac{m_{d^B}}{m_{d^A}} = \frac{v^B}{v^A}\sqrt{\frac{1 + (\tan\beta^A)^2}{1 + (\tan\beta^A)^2\left(\frac{\tan\beta^B}{\tan\beta^A}\right)^2}} 
     \stackrel{\frac{\tan\beta^B}{\tan\beta^A} \to \infty}{=} 0,
\end{equation}
and
\begin{equation}\label{eq:mutanbeta}
    \frac{m_{u^B}}{m_{u^A}} = \frac{v^B}{v^A}\frac{\tan\beta^B}{\tan\beta^A}\sqrt{\frac{1 + (\tan\beta^A)^2}{1 + (\tan\beta^A)^2\left(\frac{\tan\beta^B}{\tan\beta^A}\right)^2}}
    \stackrel{\frac{\tan\beta^B}{\tan\beta^A} \to \infty}{=} \frac{v^B}{v^A} \frac{\sqrt{1 + (\tan\beta^A)^2}}{\tan\beta^A} = \frac{v^B}{v^A} \frac{1}{\sin\beta^A},
\end{equation}
where the limits are taken with the other parameters of Eq.~\eqref{eq:Basis} kept fixed. What these equations imply is that there is an upper limit on the difference between the masses of the mirror up and mirror down of $m_{u^A} v^B/(v^A\sin\beta^A)$. Since increasing this mass difference increases the splitting between the mirror proton and mirror neutron, a lower $\tan\beta^A$ leads to a potentially lower amount of mirror atoms. On top of that, it will be shown in the next section that a lower $\tan\beta^A$ can accommodate a smaller $v^B/v^A$, which as already stated benefits the model. The only drawback is that obtaining a large $\tan\beta^B/\tan\beta^A$ requires more tuning at low $\tan\beta^A$, albeit this will be shown not to be too drastic.

\section{Collider and flavour constraints}\label{Sec:Constraints}
This section discusses how the different collider and flavour constraints are applied and which regions of parameter space they restrict. The tuning is also defined and limits on the parameters of Eq.~\eqref{eq:Basis} are presented. The constraints discussed in this section are common for either Two Higgs Doublet Models or the Twin Higgs and their treatment is fairly standard up to small modifications. Note that some of the experimental inputs, especially the collider ones, are fairly recent. This results in the constraints on certain Twin Higgs aspects being noticeably stronger than certain limits from the literature for which there has not been any update in quite some time.

\subsection{Higgs signal strengths}\label{sSec:HSS}
Obviously, the fact that the Higgs boson in the MT2HDM is a linear combination of different scalars leads to deviations of the Higgs signal strengths from their Standard Model values, which constrains the model. A convenient approach to study these observables is the $\kappa$-framework \cite{Heinemeyer:2013tqa}. Given a production mechanism $i$ with cross section $\sigma_i$ or decay process $i$ with width $\Gamma_i$, the parameters $\kappa_i$'s are defined such that 
\begin{equation}\label{eq:Defkappa}
  \kappa_i^2 = \frac{\sigma_i}{\sigma_i^{\text{SM}}} \quad \text{or} \quad \kappa_i^2 = \frac{\Gamma_i}{\Gamma_i^{\text{SM}}},
\end{equation}
where $\sigma_i^{\text{SM}}$ and $\Gamma_i^{\text{SM}}$ are the corresponding SM quantities. At leading order, $\kappa_i$ simply corresponds to the ratio of the Higgs coupling to particle $i$ and its SM value. For the bottom quark, this corresponds to
\begin{equation}\label{eq:Ybbeff}
  \kappa_b = \frac{v^A}{v_1^A} R^e_{11}.
\end{equation}
Similar results hold for the strange and down quarks. For the top quark, the equivalent result is
\begin{equation}\label{eq:Ytteff}
  \kappa_t = \frac{v^A}{v_2^A} R^e_{12},
\end{equation}
with similar results for the charm and up quarks. Lepton couplings deviate from their Standard Model values in a similar fashion to Eq.~\eqref{eq:Ybbeff} for Type II and like Eq.~\eqref{eq:Ytteff} for Type Y. The couplings of the Higgs to a pair of $W$ or $Z$ bosons deviate from their SM value by a factor of
\begin{equation}\label{eq:YWZeff}
  \kappa_W = \kappa_Z = \frac{R^e_{11} v_1^A + R^e_{12} v_2^A}{v^A}.
\end{equation}
Eqs.~\eqref{eq:Ybbeff} to \eqref{eq:YWZeff} can be shown to reduce to the equivalent Two Higgs Doublet Model results in the proper limit. This will be the case for all results in this paper.

In addition to deviations in its couplings, the Higgs can decay to mirror particles and thus lead to invisible decays. The decay width to mirror bottom quarks is
\begin{equation}\label{eq:Gammabbbarmirror}
  \Gamma_{\bar{b}^B b^B}^{h_1} = \frac{N_c}{8\pi}\left(\frac{m_{b^A} R^e_{13}}{v_{1}^A}\right)^2 \frac{\left(m_{h_1}^2 -4 m_{b^B}^2\right)^{3/2}}{m_{h_1}^2},
\end{equation}
where $N_c$ is the number of colours, i.e. 3 for quarks and 1 for leptons. Similar results hold for the mirror strange and down quarks. The decay width to mirror charms is given by
\begin{equation}\label{eq:Gammaccbarmirror}
  \Gamma_{\bar{c}^B c^B}^{h_1} = \frac{N_c}{8\pi}\left(\frac{m_{c^A} R^e_{14}}{v_{2}^A}\right)^2 \frac{\left(m_{h_1}^2 -4 m_{c^B}^2\right)^{3/2}}{m_{h_1}^2}.
\end{equation}
A similar result holds for the mirror up quark. The decay width to mirror leptons is of the form of Eq.~\eqref{eq:Gammabbbarmirror} for Type II and of Eq.~\eqref{eq:Gammaccbarmirror} for Type Y. The decay width to mirror gluons is given by a simple adaptation of the well-known SM result \cite{PhysRevD.22.178}, giving
\begin{equation}\label{eq:GammaGluonGluonmirror}
  \Gamma_{\bar{g}^B g^B}^{h_1} = \frac{(\alpha_S^B)^2 m_{h_1}^3}{128\pi^3} \left|\sum_{i\in \{d^B, s^B, b^B\}}\frac{R^e_{13}}{v_1^B}F\left(\frac{4 m_i^2}{m_{h_1}^2}\right) + \sum_{i\in \{u^B, c^B, t^B\}}\frac{R^e_{14}}{v_2^B}F\left(\frac{4 m_i^2}{m_{h_1}^2}\right)\right|^2,
\end{equation}
where
\begin{equation}\label{eq:FFunction}
  F(\tau) = -2\tau(1 + (1 - \tau)f(\tau)),
\end{equation}
with
\begin{equation}\label{eq:dFunction}
  f(\tau) = \Bigg\{ \begin{tabular}{cc} $\arcsin^2 \sqrt{\frac{1}{\tau}}$ & if $\tau \geq 1$,\\ $-\frac{1}{4}\left[\ln\left(\frac{1 + \sqrt{1 - \tau}}{1 - \sqrt{1 - \tau}}\right) - i\pi\right]^2$ & if $\tau < 1$. \end{tabular}
\end{equation}
and $\alpha_S^B$ is the mirror strong coupling obtained via one-loop running and requiring that it equals its SM equivalent at large enough scale. As is well-known and will be confirmed later, having the Higgs pass the signal strengths constraints typically requires $v^B/v^A$ to be at least $\sim 3$. This prevents the decay of the Higgs to two $W^B$'s or $Z^B$'s with one being on-shell. As such, these decays can be neglected. All other decays to mirror particles can also be neglected.

The results of Eqs.~\eqref{eq:Ybbeff} to~\eqref{eq:dFunction} can be used in combination with Ref.~\cite{Aad:2019mbh} by ATLAS and Ref.~\cite{CMS-PAS-HIG-19-005} by CMS to constrain the signal strengths. These studies consist of the most up-to-date available global fit of the signal strengths by each collaboration. Ref.~\cite{Aad:2019mbh} uses up to 80~$\text{fb}^{-1}$ of 13 TeV data and Ref.~\cite{CMS-PAS-HIG-19-005} uses up to 137~$\text{fb}^{-1}$ at also 13~TeV. These references conveniently contain all the information necessary to compute all signal strengths in terms of the $\kappa_i$'s, up to simple corrections for invisible decays. These can then be combined with the  measurements, uncertainties and correlations to compute the $\chi^2$ of the model. Our limits are obtained by combining the results of both collaborations assuming no correlations between them. A $\chi^2$ fit is performed and, when constraining a subset of parameters, the other parameters are marginalized over. These searches do not constrain directly the branching ratio to invisible, but only indirectly via the reduction of the other signal strengths. We find that the points of parameter space compatible with the Higgs signal strengths measurements typically have a branching ratio to invisible far below current constraints (see e.g. Refs.~\cite{Aaboud:2019rtt, Sirunyan:2018owy}). This fact was already noted for the Twin MSSM in Ref.~\cite{Craig:2013fga}. We therefore do not apply a constraint on the branching ratio to invisible directly. The constraints presented are at 95$\%$ confidence level.

The first consequence of the Higgs signal strengths constraints is that they force $v^B/v^A$ to be considerably larger than one. This requires some adjustment of the soft masses against terms coming from the $SU(4)$ breaking and is sometimes known as the irreducible tuning of Twin Higgs. It is typically quoted as being of $\mathcal{O}(20\%)$. The second consequence is a restriction, for relatively low $v^B/v^A$, on the allowed values of $\tan\beta^B/\tan\beta^A$ and $\tan\beta^A$. This will be discussed more carefully in Sec.~\ref{sSec:Limits}.

\subsection{Flavour physics}\label{sSec:Flavour}
This subsection explains how the different constraints from flavour physics are applied. Only the most relevant observables are included. These constraints are standard to Two Higgs Doublet Models and the resulting bounds are mostly unchanged.

\subsubsection*{$B \to X_s \gamma$}
The branching ratios of B mesons to final states containing a strange quark and a photon are well-known to be one of the most important constraints on a charged Higgs (see e.g. Ref.~\cite{Haller:2018nnx}). As the charged Higgs of the $A$ sector is prevented from mixing with that of the $B$ sector, the standard Two Higgs Doublet Model computations can be applied directly. To do this, the contributions of the charged Higgs are computed using the formulae of Ref.~\cite{Enomoto:2015wbn} with both the LO and NLO contributions. The SM prediction and its theoretical uncertainty are also taken from that paper. The experimental measurement is taken from Ref.~\cite{Amhis:2019ckw}. The total uncertainty is computed by taking the sum in quadrature of the experimental and theoretical uncertainties. A point of parameter space is then considered excluded if the prediction for the branching ratio $B \to X_s \gamma$ is outside the two-sided 95$\%$ confidence interval around the experimental measurement.

This flavour constraint effectively puts a lower limit on the mass of the charged Higgs of $\sim 600$~GeV for both Type II and Type Y. This lower limit is fairly constant for $\tan\beta^A$ larger than a few, but increases drastically as $\tan\beta^A$ approaches 0.

\subsubsection*{$B_q \to \mu^+\mu^-$}
The branching ratios of B mesons to leptons also provide important constraints on Two Higgs Doublet Models. To translate these constraints to the MT2HDM, we use the analytical results of Ref.~\cite{Enomoto:2015wbn}. Strictly speaking, these results do not apply directly, but should work to more than sufficient accuracy for the following reasons. At low $\tan\beta^A$, the new contributions to these branching ratios are dominated by the charged Higgs. Since it does not mix with its mirror partner, the results remain the same. At large $\tan\beta^A$, neutral Higgses can make sizable contributions. Since these can mix with the mirror scalars, there will in general be deviations from the Two Higgs Doublet Model results as some of the couplings will be modified. For the 125~GeV Higgs, Higgs signal strengths constraints force the couplings to be very close to their SM values and this effect is therefore negligible. The contribution of the heavier CP-even scalars are simply negligible. For the range of values we will consider, the most $A$-like non-Goldstone CP-odd scalar is almost purely $A$-like and the contribution of the other CP-odd scalar is consequently negligible. Both the decays $B_s \to \mu^+\mu^-$ and $B_d \to \mu^+\mu^-$ are taken into account. Their theoretical predictions are taken from Ref.~\cite{Enomoto:2015wbn} and their experimental values from Ref.~\cite{Amhis:2019ckw} and~\cite{Tanabashi:2018oca} respectively. The statistical treatment is the same as for $B \to X_s \gamma$.

The effect of these constraints is twofold. First, they impose some lower limit on $\tan\beta^A$ for both Type II and Type Y. Second, they impose an upper limit on $\tan\beta^A$ for Type II but not for Type Y.

\subsubsection*{$\Delta m_q$}
The last flavour observables considered are the oscillation frequencies of B mesons $\Delta m_q$. Both $\Delta m_d$ and $\Delta m_s$ are taken into account. The Two Higgs Doublet Model contributions to these observables are again dominated by the charged Higgs and the standard results can therefore be reused in the MT2HDM. The additional contributions from the charged Higgs as well as the SM predictions are taken from Ref.~\cite{Enomoto:2015wbn}. The experimental measurements and their uncertainties are taken from Ref.~\cite{Amhis:2019ckw}. The statistical treatment is the same as for the other flavour observables.

These flavour observables impose a lower bound on $\tan\beta^A$ that decreases as the mass of the charged Higgs is increased. This bound applies equally to both Type II and Type~Y.

\subsection{Collider searches}\label{sSec:Collider}
This section presents how the different collider constraints are applied. The searches selected are those that either constrain region unconstrained by flavour physics or at least compete with them.

\subsubsection*{$H^+ \to t\overline{b}$}
One of the most common search strategies for a charged Higgs is via its decay to a top and a bottom. Some basic modifications must however be performed to apply these bounds to the MT2HDM.

The charged Higgs of the $A$ sector can decay via several channels. First, it can decay to a top and bottom quark with a decay width of
\begin{equation}\label{eq:ChargedHiggsDecaytb1}
  \Gamma_{\bar{b}^A t^A}^{H^+_p} = \frac{N_c}{8\pi}\left[(A_L^2 + A_R^2)(m_{H^+_p}^2 - m_{t^A}^2 - m_{b^A}^2) - 4A_R A_L m_{t^A}m_{b^A}\right]\frac{|\vec{p}_{t^A}|}{m_{H^+_p}^2},
\end{equation}
with
\begin{equation}\label{eq:ChargedHiggsDecaytb2}
   A_L = -\frac{\sqrt{2}m_{t^A} }{v_2^A}R^+_{p1}, \qquad
   A_R =  \frac{\sqrt{2}m_{b^A} }{v_1^A}R^+_{p2},
\end{equation}
where $p$ is the index corresponding to the SM charged Higgs (3 or 4), $|\vec{p}_{t^A}|$ is the norm of the center of mass momentum of the top quark, which is easily computed in terms of standard kinematics, and the CKM matrix is approximated as diagonal. Similar results hold for first and second generation quarks. The decay width to a tau and a neutrino is of the same form as Eq.~\eqref{eq:ChargedHiggsDecaytb1}, but with
\begin{equation}\label{eq:ChargedHiggsDecaytaunu}
  \begin{aligned}
    & \text{Type II:} & A_L = 0, \quad &  A_R = \frac{\sqrt{2}m_{\tau^A} }{v_1^A}R^+_{p1},\\
    & \text{Type Y: } & A_L = 0, \quad &  A_R = \frac{\sqrt{2}m_{\tau^A} }{v_2^A}R^+_{p2}.
  \end{aligned}
\end{equation}
Similar results hold for first and second generation leptons. The decay to a $W^A$ and a CP-even scalar $h_i$ is also possible and has a width of
\begin{equation}\label{eq:CheargedHiggsDecayWscalar}
  \Gamma_{W^A h_i}^{H^+_p} = \frac{G_F^A}{\sqrt{2}\pi}(R^+_{p1}R^e_{i1} + R^+_{p2}R^e_{i2})^2|\vec{p}_{h_i}|^3,
\end{equation}
where $G_F^A$ is the SM Fermi constant and the decay is assumed kinematically allowed. A similar result holds for the decay to a CP-odd scalar, but with $R^e$ replaced by $R^o$.

With these results, it is simple to apply bounds from LHC searches. For a charged Higgs heavier than the top, the dominant production mechanisms are $p p \to \bar{t} H^+ + X$ and its conjugate process. These were studied in Refs.~\cite{Aaboud:2018cwk, CMS-PAS-HIG-18-004, Sirunyan:2020hwv}, with all of them using approximately 36~$\text{fb}^{-1}$ of integrated luminosity at 13~TeV. This was done in the mono and di-lepton channels for Refs.~\cite{Aaboud:2018cwk, CMS-PAS-HIG-18-004} and the all-jet final state in Ref.~\cite{Sirunyan:2020hwv}. The latter channel leads to much weaker limits and is not taken into account in our analysis. The cross section with NLO QCD corrections is unchanged from the standard Two Higgs Doublet Model case and is taken from Refs.~\cite{Degrande:2015vpa, Flechl:2014wfa, deFlorian:2016spz, Dittmaier:2009np, Berger:2003sm}. It is computed by combining the four-flavour scheme (4fs) and five-flavour scheme (5fs) using the Santander matching \cite{Harlander:2011aa}. The branching ratio to a top and a bottom is computed using Eqs.~\eqref{eq:ChargedHiggsDecaytb1} to~\eqref{eq:CheargedHiggsDecayWscalar}. The resulting cross section times branching ratio can then be compared with the limits of Refs.~\cite{Aaboud:2018cwk, CMS-PAS-HIG-18-004}. A point of parameter space is considered excluded if the cross section times branching ratio is above any individual 95\% CLs limit from one of these searches \cite{Read:2002hq, Junk:1999kv}.

This constraint imposes both an upper and lower limit on $\tan\beta^A$. It applies to Type II and Type Y. In principle, the modification of the decay widths with respect to their standard Two Higgs Doublet Model counterparts changes the limits on $\tan\beta^A$ for a given mass of the charged Higgs. In practice, the branching ratio to a top and a bottom quark is always very close to 1 in both cases and the limits are almost identical.

\subsubsection*{$H/A \to \tau\bar{\tau}$}
Another important constraint is the production of CP-even or CP-odd heavy scalars that decay to a pair of taus. Since these generally include an admixture of mirror scalars, the treatment is modified from the Two Higgs Doublet Model.

Neutral scalars can decay to a pair of fermions. Assuming the channels are open, the main decays of this type have the following widths
\begin{equation}\label{eq:CPEvenOddFF}
  \begin{aligned}
    & \Gamma_{\bar{t}^A t^A}^{h_i} = \frac{N_c}{8\pi}\left(\frac{m_{t^A} R^e_{i2}}{v_{2}^A}\right)^2 \frac{\left(m_{h_i}^2 -4 m_{t^A}^2\right)^{3/2}}{m_{h_i}^2}, &
    & \Gamma_{\bar{b}^A b^A}^{h_i} = \frac{N_c}{8\pi}\left(\frac{m_{b^A} R^e_{i1}}{v_{1}^A}\right)^2 \frac{\left(m_{h_i}^2 -4 m_{b^A}^2\right)^{3/2}}{m_{h_i}^2}, \\
    & \Gamma_{\bar{t}^A t^A}^{a_i} = \frac{N_c}{8\pi}\left(\frac{m_{t^A} R^o_{i2}}{v_{2}^A}\right)^2 \sqrt{m_{a_i}^2 -4 m_{t^A}^2},                               &
    & \Gamma_{\bar{b}^A b^A}^{a_i} = \frac{N_c}{8\pi}\left(\frac{m_{b^A} R^o_{i1}}{v_{1}^A}\right)^2 \sqrt{m_{a_i}^2 -4 m_{b^A}^2}.
  \end{aligned}
\end{equation}
Similar expressions hold for the other generations. The decay widths to taus are of the same form as to bottoms for Type II and as to tops for Type Y. If kinematically allowed, the CP-even scalars can decay to massive gauge bosons with widths
\begin{equation}\label{eq:CPEvenBB}
  \begin{aligned}
    & \Gamma_{W^A W^A}^{h_i} = \frac{(G_F^A)^2}{8\pi}(v_2^A R^e_{i2} + v_1^A R^e_{i1})^2 \frac{(12m_{W^A}^4 - 4m_{W^A}^2 m_{h_i}^2 + m_{h_i}^4)}{m_{h_i}^2}\sqrt{m_{h_i}^2 -4 m_{W^A}^2},\\
    & \Gamma_{Z^A Z^A}^{h_i} = \frac{(G_F^A)^2}{16\pi}(v_2^A R^e_{i2} + v_1^A R^e_{i1})^2 \frac{(12m_{Z^A}^4 - 4m_{Z^A}^2 m_{h_i}^2 + m_{h_i}^4)}{ m_{h_i}^2}\sqrt{m_{h_i}^2 -4 m_{Z^A}^2}.
  \end{aligned}
\end{equation}
If kinematically allowed, CP-even and CP-odd scalars can also decay to a $Z$ boson and a scalar of opposite parity with widths
\begin{equation}\label{eq:CPEvenOddZOddEven}
  \begin{aligned}
  & \Gamma_{a^j Z^A}^{h_i} =  \frac{G_F^A}{\sqrt{2}\pi}(R^e_{i1}R^o_{j1} + R^e_{i2}R^o_{j2})^2|\vec{p}_{a_j}|^3, \\
  & \Gamma_{h^i Z^A}^{a_j} =  \frac{G_F^A}{\sqrt{2}\pi}(R^e_{i1}R^o_{j1} + R^e_{i2}R^o_{j2})^2|\vec{p}_{h_i}|^3.
  \end{aligned}
\end{equation}
The generalization of Eqs.~\eqref{eq:CPEvenOddFF} to~\eqref{eq:CPEvenOddZOddEven} to mirror particles is trivial. The decay of $h_i$ to a pair of $h_1$ is taken into account numerically due to the complexity of the result, albeit this decay width is generally rather small.

These results can then be used in conjunction with the latest search for a scalar or pseudo-scalar decaying to a pair of taus found in Ref.~\cite{Aad:2020zxo}. This search by ATLAS uses 139~$\text{fb}^{-1}$ of integrated luminosity at 13~TeV. We concentrate on b-associated production. The corresponding cross section is taken from Ref.~\cite{deFlorian:2016spz}, which is computed using \texttt{SUSHI} \cite{Harlander:2012pb, Harlander:2016hcx}, and rescaled according to the recommendations of Ref.~\cite{Harlander:2012pb}. Production via gluon fusion is also considered in the search. We implemented it using the leading order cross section and found limits considerably inferior to those from b-associated production. The reason is that the cross section for gluon fusion decreases with respect to that of b-associated production as $\tan\beta^A$ increases and, in the Two Higgs Doublet Model, is relatively small around the upper limit on $\tan\beta^A$. This is accentuated in the MT2HDM, as significant branching ratios of the heavier Higgses to invisible push up this limit. Gluon fusion is therefore not taken into account in the final results. The branching ratios are computed using Eqs.~\eqref{eq:CPEvenOddFF} to~\eqref{eq:CPEvenOddZOddEven}.  Ref.~\cite{Aad:2020zxo} imposes limits by summing the cross section times branching ratio for the heavy CP-even scalar and the CP-odd scalar. This is justified for the MSSM, as current constraints force their mass splitting to be below detector resolution and they form an almost degenerate pair. In the MT2HDM, there are instead two such pairs and a single heavy CP-even Higgs. For each pair, the cross sections times branching ratios of the CP-even and CP-odd scalars are summed. The singlet is treated individually. The three resulting quantities are then compared to the 95\% CLs limits from Ref.~\cite{Aad:2020zxo} and a point of parameter space is considered excluded if any of them individually surpasses its corresponding limit.

The effect of this collider search is to impose an upper limit on $\tan\beta^A$ for Type II. This turns out to be by far the strongest upper limit on $\tan\beta^A$ for this type. We do mention that the presence of the mirror sector tends to decrease both the cross sections for the mostly $A$-like scalars and their branching ratio to $\tau\bar{\tau}$ with respect to the Two Higgs Doublet Model. This weakens the constraints on individual scalars. However, the singlet mentioned in the previous paragraph is not something that exists in the standard Two Higgs Doublet Model and can in principle be more constrained than the mostly $A$-like scalars. Which effect wins depends non-trivially on many factors.

\subsubsection*{$H/A \to b\bar{b}$}
The last collider observable considered is the production of neutral scalars decaying to a pair of bottom quarks in the b-associated channel. Experimental constraints are taken from Ref.~\cite{Aad:2019zwb} by ATLAS which uses 28~$\text{fb}^{-1}$ of integrated luminosity at 13~TeV. The branching ratios and cross section are computed as for $H/A \to \tau\bar{\tau}$. A point of parameter space is considered excluded if its cross section times branching ratio is above the 95\% CLs limit of Ref.~\cite{Aad:2019zwb}.

The effect of this constraint is to impose an upper limit on $\tan\beta^A$ for both Type II and Type Y. This happens to be the most stringent upper limit on $\tan\beta^A$ for Type Y. The discussion about how the bounds are modified with respect to the Two Higgs Doublet Model for $\tau\bar{\tau}$ also applies for $b\bar{b}$.

\subsection{Tuning and other considerations}\label{sSec:Tuning}
As is the case with most models of physics beyond the Standard Model, the Twin Higgs is subject to a certain amount of tuning. We discuss three possible sources.

First, obtaining a Higgs boson whose properties are sufficiently close to those of the Standard Model requires $v^B/v^A$ to be larger than a factor of roughly 3. This typically requires making $\theta_2$ small, which must be done by adjusting the soft masses. We will adopt a definition for this tuning similar to that of Ref.~\cite{Craig:2013fga} via
\begin{equation}\label{eq:Tuningvonf}
  \Delta_{v/f} = \max_{p\in \mathcal{P}} \left|\frac{\partial \ln \left((v^A)^2/f^2\right)}{\partial \ln p}\right|,
\end{equation}
where $f^2=(v^A)^2 + (v^B)^2$ and $\mathcal{P}$ is the set of parameters of the potential of Eq.~\eqref{eq:GeneralPotential1}.\footnote{For parameter such as $\Delta m_2^2$, it is understood that the square is part of the parameter, i.e. the derivative is with respect to $\Delta m_2^2$.} The corresponding tuning is $T_{\theta_2}=\Delta_{v/f}^{-1}$.

Second, obtaining a large $\tan\beta^B/\tan\beta^A$ will sometimes require making $v_1^B/v_1$ small, i.e. making $\theta_1$ close to $\pi/2$. A similar tuning can then be defined via
\begin{equation}\label{eq:Tuningv1Aonf1}
  \Delta_{v_1^B/v_1} = \max_{p\in \mathcal{P}} \left|\frac{\partial \ln \left((v_1^B)^2/v_1^2\right)}{\partial \ln p}\right|.
\end{equation}
The corresponding tuning is $T_{\theta_1}=\Delta_{v_1^B/v_1}^{-1}$.

Third, keeping $f$ relatively small can require some adjusting and a resulting tuning could be defined via
\begin{equation}\label{eq:Tuning}
  \Delta_{f} = \max_{p\in \mathcal{P}} \left|\frac{\partial \ln f^2}{\partial \ln p}\right|.
\end{equation}
The corresponding tuning would be $\Delta_{f}^{-1}$. In practice, the complication with this term is that it is highly dependent on the UV completion. For example, the earliest SUSY UV completions tried to generate an $SU(4)$ preserving quartic via an F-term and a singlet chiral superfield with a large soft mass, resulting in a total tuning of $\mathcal{O}(1\%)$ \cite{Craig:2013fga}. More modern UV completions where the $SU(4)$ preserving quartic is instead generated via the D-term potential of a new gauge group (Abelian or not) can bring the total tuning to the $\mathcal{O}(10\%)$ level \cite{Badziak:2017syq, Badziak:2017kjk, Badziak:2017wxn}. In principle, an appropriate UV completion might be able to bring $\Delta_{f}$ to a value not so different from 1, in which case it would not really be a tuning.

To remain as generic as possible, the rest of the paper will generally focus on $T_{\theta_1}$ and $T_{\theta_2}$. Information as to how the results can be embedded into different UV completions will be discussed in the conclusion.

\subsection{Limits on vevs}\label{sSec:Limits}
We finish this section by assembling the different constraints to obtain limits on the parameters of Eq.~\eqref{eq:Basis}.

To excellent approximation, the cosmology only depends on the potential via the three parameters of Eq.~\eqref{eq:Basis}. Which values of these parameters are allowed however depends on the full parameter space, which is 16 dimensional. In addition, two constraints exist in that the mass of the Z and Higgs bosons must be properly reproduced.  As such, the strategy we adopt is to project the allowed region of the full parameter space on the cosmologically relevant parameter space. This is done via Monte Carlo simulation.

The figures we present use $\tan\beta^A$ and $\tan\beta^B/\tan\beta^A$ as axes with $v^B/v^A$ kept to different fixed values. The parameters are sampled uniformly in the range:
\begin{equation}\label{eq:ScanRange}
  \begin{tabular}{llll}
    $\frac{\mu_1^2}{\mu_2^2} \in [-64,\; 64]$, & $\frac{B_\mu}{\mu_2^2} \in [0,\; 1]$, & $\frac{\Delta m_1^2}{\mu_2^2} \in [-100,\; 0]$, & $\frac{\Delta B_\mu}{\mu_2^2} \in [0,\; 0.1]$, \\
    $\lambda_1 \in [1,\; 2]$,                  & $\lambda_2 \in [1,\; 2]$              & $\alpha_1 \in [-0.5,\; 0]$,                                                                    \\
    $\lambda_3 \in [-0.2,\; 0.2]$,             & $\lambda_4 \in [-0.2,\; 0.2]$,        & $\lambda_5 \in [-0.2,\; 0]$,                                                                   \\
    $\beta_3 \in [-0.2,\; 0.2]$,               & $\beta_4 \in [-0.2,\; 0.2]$,          & $\beta_5 \in [-0.2,\; 0]$.
  \end{tabular}
\end{equation}
The parameters which are not specified ($\mu_2^2$, $\Delta m_2^2$ and $\alpha_2$) are adjusted to reproduce the correct $v^B/v^A$, $m_{Z^A}$ and $m_{h_1}$. This scan is of course not exhaustive, but will prove to be sufficient to demonstrate that mirror neutron dark matter is indeed possible. With the exception of $\Delta m_1^2$, parameters that break the $SU(4)$ symmetry are maintained deliberately small. The parameters $\lambda_1$ and $\lambda_2$ are kept relatively large, as this usually minimizes $\Delta_{f}$ in UV completions \cite{Craig:2013fga}, but not so large as to cause perturbativity problems. The other $SU(4)$ preserving parameters could also in principle be taken as large, but the values considered will prove to be sufficient. The range of the other parameters is chosen to properly sample the $\tan\beta^A$ and $\tan\beta^B/\tan\beta^A$ space. In addition to the previous constraints, a point is not taken into account if the potential is not bounded from below or in the presence of charge/CP breaking vacua in either sectors. The chosen range of parameter space prevents this from being a common occurrence.

\begin{figure}[t!]
  \centering
   \captionsetup{justification=centering}
    \begin{subfigure}{0.495\textwidth}
    \centering
    \caption{$T_{\theta_2}$, $v^B/v^A = 2.5$}
    \includegraphics[width=\textwidth]{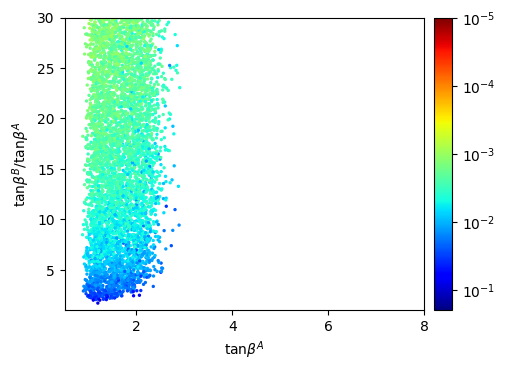}
    \label{fig:Fig1_250_t1}
  \end{subfigure}
  \begin{subfigure}{0.495\textwidth}
    \centering
    \caption{$T_{\theta_1}$, $v^B/v^A = 2.5$}
    \includegraphics[width=\textwidth]{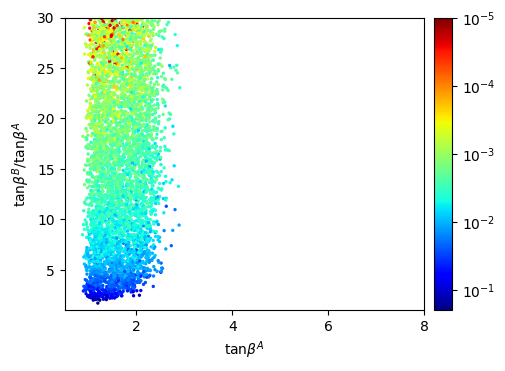}
    \label{fig:Fig1_250_t2}
  \end{subfigure}
  \vspace{0.0cm}
  \begin{subfigure}{0.495\textwidth}
    \centering
    \caption{$T_{\theta_2}$, $v^B/v^A = 3.0$}
    \includegraphics[width=\textwidth]{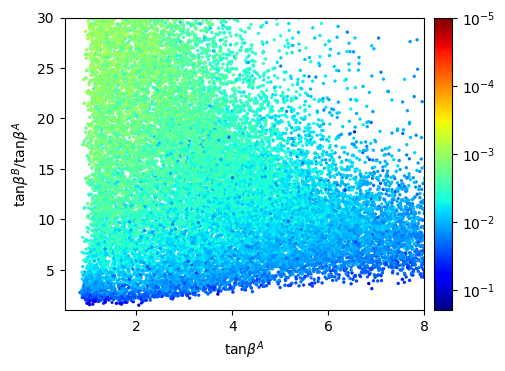}
    \label{fig:Fig1_300_t1}
  \end{subfigure}
  \begin{subfigure}{0.495\textwidth}
    \centering
    \caption{$T_{\theta_1}$, $v^B/v^A = 3.0$}
    \includegraphics[width=\textwidth]{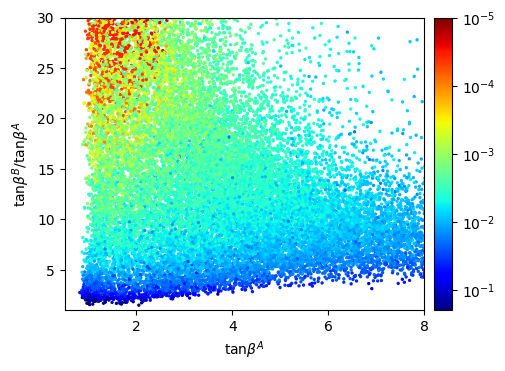}
    \label{fig:Fig1_300_t2}
  \end{subfigure}
  \vspace{0.0cm}
  \begin{subfigure}{0.495\textwidth}
    \centering
    \caption{$T_{\theta_2}$, $v^B/v^A = 3.5$}
    \includegraphics[width=\textwidth]{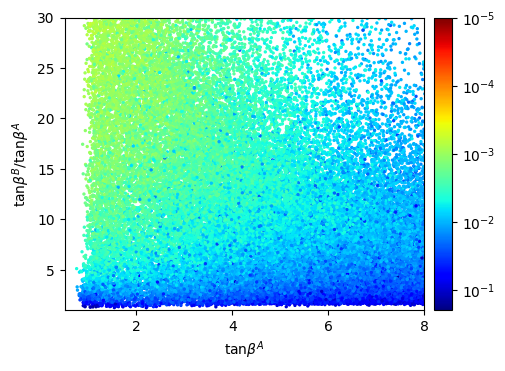}
    \label{fig:Fig1_350_t1}
  \end{subfigure}
  \begin{subfigure}{0.495\textwidth}
    \centering
    \caption{$T_{\theta_1}$, $v^B/v^A = 3.5$}
    \includegraphics[width=\textwidth]{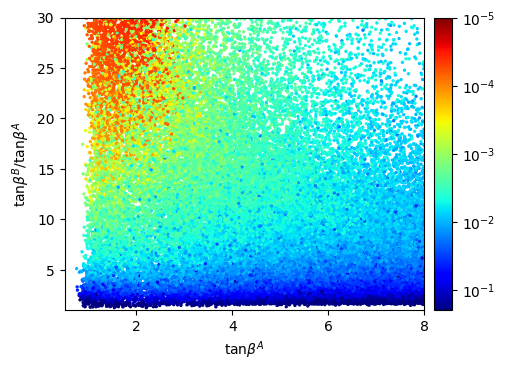}
    \label{fig:Fig1_350_t2}
  \end{subfigure}
  \captionsetup{justification=justified}
\caption{Values of $\tan\beta^A$ and $\tan\beta^B/\tan\beta^A$ of the unexcluded points for different fixed $v^B/v^A$ and Type II. The colour corresponds to $T_{\theta_2}$ (left) and $T_{\theta_1}$ (right).}\label{fig:ScatterConstraints}
\end{figure}

A few example plots are shown for Type II in Fig.~\ref{fig:ScatterConstraints}, using $4\times 10^5$ points as in all other plots.
Constraints for Type Y will be presented in Sec.~\ref{sSec:Results}. They are not so different from those for Type II and we do not include them here to avoid unnecessary redundancy. The colour of the points corresponds to $T_{\theta_2}$ in Figs.~\ref{fig:Fig1_250_t1}, \ref{fig:Fig1_300_t1}, \ref{fig:Fig1_350_t1} and to $T_{\theta_1}$ in Fig.~\ref{fig:Fig1_250_t2}, \ref{fig:Fig1_300_t2}, \ref{fig:Fig1_350_t2}. As can be seen, a very large $\tan\beta^B/\tan\beta^A$ is indeed possible and over a large range of $\tan\beta^A$. This typically requires more tuning at low $\tan\beta^A$, as the region of small $\tan\beta^A$ and large $\tan\beta^B/\tan\beta^A$ corresponds to a very small $\theta_2$. Larger values of $\tan\beta^A$ than what is shown in the plots are possible, but the cosmology will reveal that there is no gain from having a $\tan\beta^A$ larger than a few. For small values of $v^B/v^A$, only the region of low $\tan\beta^A$ and/or large $\tan\beta^B/\tan\beta^A$ is allowed. This is again because this region corresponds to a smaller $\theta_2$, which allows for a more $A$-like $h_1$ and therefore for the possibility of the Higgs signal strengths requirements being met. We do mention that the inclusion of the second Higgs doublet and its mirror copy allows for smaller values of $v^B/v^A$ than would be permitted in the standard Mirror Twin Higgs. The low number of points in the upper right corner, when this region is allowed, is more of a numerical artefact as $v_2^B$ is so small there as to be difficult to estimate numerically. The best fit of the $\chi^2$ of the Higgs signal strengths is 21.93 for Type II and 22.04 for Type Y. These are either the same value as for the Standard Model (22.04) or a slightly lower one. This is unsurprising, as the signal strengths are generally scaled almost uniformly by a factor less than one and the experimental measurements prefer a uniform scaling by a factor slightly larger than one. In the end, the flavour and collider searches constraints cut relatively few points and only at low or large $\tan\beta^A$. This was to be expected, as the Higgs is a pseudo-Goldstone boson and the other scalars are typically much heavier.

As is clear from the figures, the two tunings $T_{\theta_1}$ and $T_{\theta_2}$ are highly correlated when both are small. This can be understood as follows. The quantity $T_{\theta_2}$ is a measure of the tuning necessary to make $\theta_2$ small and $T_{\theta_1}$ the one necessary to make $\theta_1$ close to $\pi/2$. On one hand, $H_1^A$ leads to an effective tadpole for $H_2^A$ via the $B_\mu$ term \cite{Beauchesne:2015lva}. A non-zero $\theta_1$ therefore leads to $\theta_2$ also being non-zero. In other words, pushing $\theta_1$ close to $\pi/2$ also ensures that $\theta_2$ is non-zero. On the other hand, this whole argument can be reversed and pushing $\theta_2$ close to zero ensures that $\theta_1$ is below $\pi/2$. All in all, it might not be wise to consider them as two independent sources of tuning. Multiplying  $T_{\theta_2}$ and $T_{\theta_1}$ together to obtain some form of total tuning might therefore not accurately reflect the required amount of adjustment of the model. For the rest of the paper, we will hence concentrate on the minimum of these two tunings.

\section{Cosmological constraints, mirror BBN and final results}\label{Sec:Cosmo}
In this section, we discuss the different constraints from cosmology, explain how mirror BBN is treated and present the final results.

\subsection{Constraints}\label{sSec:Constraints}
We consider two main constraints from cosmology. The first one is on the number of effective extra relativistic degrees of freedom $N_{\text{eff}}$, which is measured by Planck to be $2.99\pm 0.17$ \cite{Aghanim:2018eyx}. This imposes an upper limit on the number of new relativistic degrees of freedom $\Delta N_{\text{eff}}$ and thus on the ratio of the temperatures of the $A$ and $B$ sectors via
\begin{equation}\label{eq:rT}
  r_T = \frac{T^B}{T^A} = \left(\frac{\Delta N_{\text{eff}}}{7.4}\right)^{1/4},
\end{equation}
where we borrowed the notation of Ref.~\cite{Chacko:2018vss}, $T^M$ is the temperature of sector $M$ and electron recombination is assumed to have aleady taken place in both sectors. In practice, this means that $r_T$ is forced to be below $\sim 0.46$ at 95\% CL.

The second constraint considered is on the fraction of dark atoms $X_{\text{DA}}$. Ref.~\cite{Fan:2013yva} makes the claim that 10\% of dark matter could have arbitrarily large self-interactions and still be compatible with all current observations. It is clear however that there is a certain amount of uncertainty on this number. In addition, Ref.~\cite{Chacko:2018vss} claims that the bound might actually be brought down to the few percent level in the not-so-distant future. As such, we will indeed take the bound to be 10\%, but we will present additional contours of $X_{\text{DA}}$ in case more precise limits become available in the future.

\subsection{Computation of the relic abundances}\label{sSec:CRA}
The computation of the different abundances is done via simple approximations. More complicated computations would in principle be possible, but some of the uncertainties are sufficiently large to overhelm any resulting gain in precision. This stems in large part from the fact that many parameters crucial to BBN can be precisely measured experimentally but not computed yet to high precision. Since properties of the dark sector are obviously not currently measured, this can result in large variations in some abundances.

Three properties of the mirror sector must be determined before proceeding to the computation of the relic abundances (see Ref.~\cite{Chacko:2018vss} for a similar treatment from which we take inspiration). First, the value of the mirror QCD scale $\Lambda_\text{QCD}^B$ is trivially computed by assuming that the strong coupling constants of the two sectors are equal at sufficiently large scale and running them down via the one-loop renormalization equation. A value of the QCD scale of the SM sector $\Lambda_\text{QCD}^A$ of 220~MeV is taken, though our cosmology results ultimately only depend on the ratio of $\Lambda_\text{QCD}^B/\Lambda_\text{QCD}^A$.

Second, the SM deuterium binding energy $B_{D^A}$ is available from lattice QCD at different pion masses $m_{\pi^A}$ \cite{Orginos:2015aya, Savage:2015eya, Beane:2011iw, Beane:2012vq, Yamazaki:2012hi, Yamazaki:2015asa}. Using the least squares method, a linear fit is performed on the different combinations $(m_{\pi^A}/\Lambda_\text{QCD}^A, B_{D^A}/\Lambda_\text{QCD}^A)$. This gives
\begin{equation}\label{eq:BD}
  \frac{B_{D^A}}{\Lambda_\text{QCD}^A} = B_1 \frac{m_{\pi^A}}{\Lambda_\text{QCD}^A} + B_2,
\end{equation}
with $B_1\approx 0.033$ and $B_2\approx -0.011$. Assuming the same relation holds in the mirror sector, the mirror binding energy $B_{D^B}$ can then be computed using the previously obtained value of $\Lambda_\text{QCD}^B$ and the mirror pion mass. The latter is computed via the Gell-Mann-Oakes-Renner formula \cite{GellMann:1968rz}, assuming the dimensionless proportionality constant is the same in both sectors. It is important to note that these estimates from lattice QCD have very large uncertainties. Since the deuterium binding energy determines when deuterium formation begins, this unfortunately means that a large uncertainty is introduced when the mirror protons decay quickly. This is one of the main reasons why the computation is limited to a simple approximation.

Third, the proton-neutron mass difference is extracted from the lattice QCD results of Fig.~3 from Ref.~\cite{Borsanyi:2014jba} (or alternatively Table~2). This result in a relation
\begin{equation}\label{eq:Deltpn}
  m_{pn}^B = m_{p^B} - m_{n^B} = C_0\left(C_1(m_{u^B} - m_{d^B}) - C_2 \alpha_{\text{EM}} \Lambda_\text{QCD}^B\right),
\end{equation}
where $\alpha_{\text{EM}}$ is the fine structure constant, $C_1\approx 0.86$, $C_2\approx 0.54$ and $C_0$ is fixed to reproduce the SM equivalent value of $m_{pn}^A$. These results are obtained by using the mass of the up and down of Ref.~\cite{Tanabashi:2018oca}. Two comments need to be made. First, we obtain a slightly larger $C_2$ than Ref.~\cite{Chacko:2018vss}. This is probably due to a different choice of $\Lambda_\text{QCD}^A$, which ultimately does not affect this result, and we reproduce rather accurately their mass splitting as given by their Eq.~(2.7). Second, the constant $C_0$ is introduced as the results of Ref.~\cite{Borsanyi:2014jba} are normalized to reproduce the experimental measurement for the physical values of $m_{u^A} - m_{d^A}$ and $\alpha_{\text{EM}}$. For the central values of these parameters, $C_0$ is one. However, we will later investigate the effects of the uncertainties on the masses of the up and down quarks. This affects the neutron-proton splitting which must then be normalized back to its experimental value.

Once these three quantities are obtained, it is trivial to compute the ratio of mirror proton and mirror neutron abundances $n_{p^B}/n_{n^B}$. We will follow mainly Refs.~\cite{Kolb:1990vq, Mukhanov:2003xs} and again Ref.~\cite{Chacko:2018vss}. At high $T^B$, mirror protons and mirror neutrons are kept in equilibrium via conversion off electrons. The corresponding rate is
\begin{equation}\label{eq:Conversion}
  \Gamma_{p^B e^B \to n^B \nu_e^B}= \frac{1 + 3g_A^2}{2\pi^3}(G_F^B)^2 (m_{pn}^B)^5 J(-\infty, -\frac{m_{e^B}}{m_{pn}^B}),
\end{equation}
where
\begin{equation}
  J(a, b) = \int_a^b\sqrt{1 - \frac{(m_{e^B}/m_{pn}^B)^2}{q^2}}\frac{q^2(q - 1)^2 dq}{\left(1 + e^{\frac{m_{pn}^B}{T_\nu^B}(q - 1)}\right)\left(1 + e^{-\frac{m_{pn}^B}{T^B}q}\right)},
\end{equation}
with $g_A = 1.27$, $G_F^B$ the mirror Fermi constant and $T_\nu^B$ the temperature of the mirror neutrinos, i.e. $T^B$ before and $(4/11)^{1/3}T^B$ after electron recombination. This process freezes-out at a temperature of the mirror sector $T_{\text{FO}^B}^B$ where this rate equals the Hubble constant $H \approx 1.66 g_\star^{1/2} (T^A)^2/m_{\text{Pl}}$, where $g_*$ is the number of relativisitc degrees of freedom and $m_{\text{Pl}}$ the Planck mass. At this point, the ratio of mirror protons and neutrons is $n_{p^B}/n_{n^B}\approx f_1 \approx \exp(-m_{pn}^B/T_{\text{FO}^B}^B)$. Assuming they are unstable, free protons will continue to decay until deuterium formation with a corresponding width of
\begin{equation}\label{eq:GammapB}
  \Gamma^{p^B}_{n^B e^B \nu_e^B} = \frac{1 + 3g_A^2}{2\pi^3}(G_F^B)^2 m_{e^B}^5\lambda_0(m_{pn}^B/m_{e^B}),
\end{equation}
where
\begin{equation}\label{eq:gamma0}
  \lambda_0(Q) = \int_1^Q dq q(q - Q)^2(q^2 - 1)^{1/2}.
\end{equation}
The temperature of the mirror sector at which the mirror deuterium bottleneck is crossed is $T^B_{\text{DB}^B} \approx (B_{D^B}/B_{D^A}) T^A_{\text{DB}^A}$, where $T^A_{\text{DB}^A}$ is taken as 0.08~MeV \cite{Mukhanov:2003xs}. This takes place at a time $t_{\text{DB}^B} = 0.301 g_\star^{-1/2} m_{\text{Pl}}r_T^2/(T^B_{\text{DB}^B})^2$, meaning that the ratio of mirror protons and mirror neutrons decreased by an additional factor $f_2 \approx \exp(-\Gamma^{p^B}_{n^B e^B \nu_e^B}t_{\text{DB}^B})$. The neutron to proton ratio at the onset of deuterium formation is then $(n_{p^B}/n_{n^B})_{\text{DB}^B} \approx f_1 \times f_2$.

After mirror deuterium formation, it is quickly converted to mirror Helium-4. In principle, mirror Helium-4 could be unstable, but in practice this is extremely unlikely in the region of parameter space of our interest. The binding energy of Helium-4 is 28.3~MeV and that of the mirror Helium-4 should in general be considerably larger (see Refs.~\cite{Yamazaki:2012hi, Yamazaki:2015asa} for lattice QCD results). There are two possible options for decays of mirror Helium-4. First, mirror Helium-4 could decay via beta decay to a positron, a neutrino and a combination of one proton and three neutrons. The most promising single beta decay would be to Hydrogen-3, a free neutron, a positron and a neutrino. However, this would require a $m_{pn}^B$ of at least $\sim 20$~MeV, which would require a very large $v^B/v^A$ and thus an unacceptable amount of tuning. Second, Helium-4 could also in principle decay via double beta decay to two positrons, two neutrinos and some combination of four neutrons. It is unclear as to whether mirror dineutrons would be stable \cite{Orginos:2015aya} and states of three or four neutrons would have binding energy of at most a few MeVs. This would require a $m_{pn}^B$ of at least $\sim 15$~MeV, which also requires unacceptable tuning. Considering the uncertainties of lattice QCD, it is still possible that mirror Helium-4 could be unstable, but it seems very unlikely at best. If it did decay, it would actually facilitate obtaining a low fraction of dark atoms and, in the very worst case scenario, our bounds can be interpreted as conservative. We will therefore assume Helium-4 to be stable from now on. The formation of heavier elements is suppressed by the absence of sufficiently stable elements of atomic mass 5 or 8, which should not change from the Standard Model barring extreme tuning. The final fraction of dark atoms $X_{\text{DA}}$ is then
\begin{equation}\label{eq:FracAtom}
  X_{\text{DA}} \approx \frac{2(n_{p^B}/n_{n^B})_{\text{DB}^B}}{1 + (n_{p^B}/n_{n^B})_{\text{DB}^B}}.
\end{equation}

\subsection{Results}\label{sSec:Results}
\begin{figure}[t!]
  \centering
   \captionsetup{justification=centering}
    \begin{subfigure}{0.495\textwidth}
    \centering
    \caption{Type II, $v^B/v^A = 2.5$}
    \includegraphics[width=\textwidth]{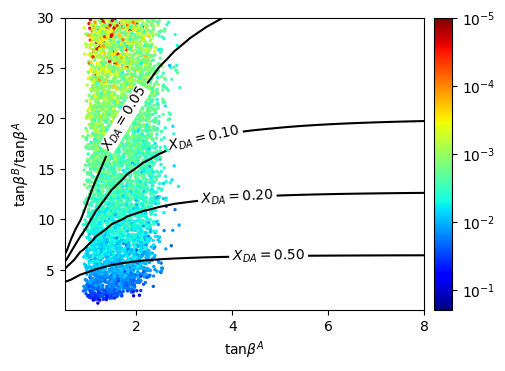}
    \label{fig:Fig2_250_II}
  \end{subfigure}
  \begin{subfigure}{0.495\textwidth}
    \centering
    \caption{Type Y, $v^B/v^A = 2.5$}
    \includegraphics[width=\textwidth]{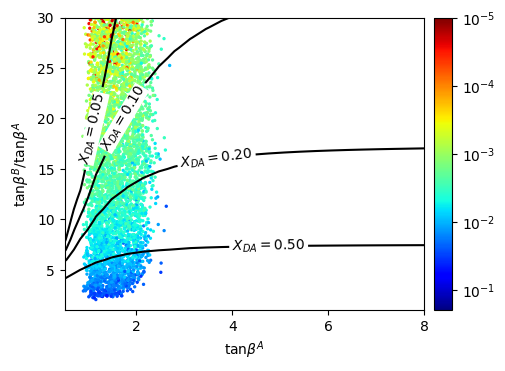}
    \label{fig:Fig2_250_Y}
  \end{subfigure}
  \vspace{0.0cm}
  \begin{subfigure}{0.495\textwidth}
    \centering
    \caption{Type II, $v^B/v^A = 3.0$}
    \includegraphics[width=\textwidth]{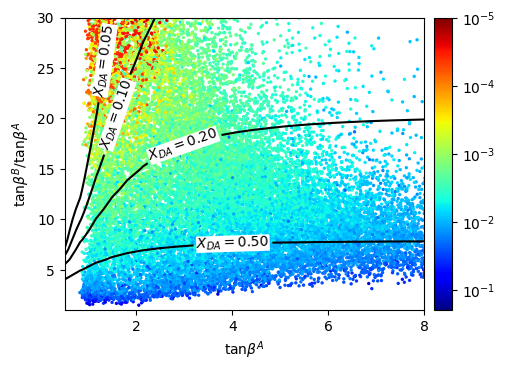}
    \label{fig:Fig2_300_II}
  \end{subfigure}
  \begin{subfigure}{0.495\textwidth}
    \centering
    \caption{Type Y, $v^B/v^A = 3.0$}
    \includegraphics[width=\textwidth]{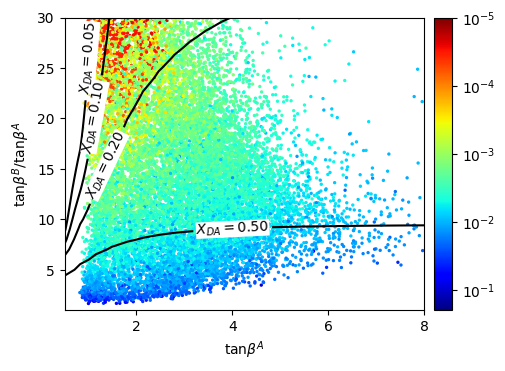}
    \label{fig:Fig2_300_Y}
  \end{subfigure}
  \vspace{0.0cm}
  \begin{subfigure}{0.495\textwidth}
    \centering
    \caption{Type II, $v^B/v^A = 3.5$}
    \includegraphics[width=\textwidth]{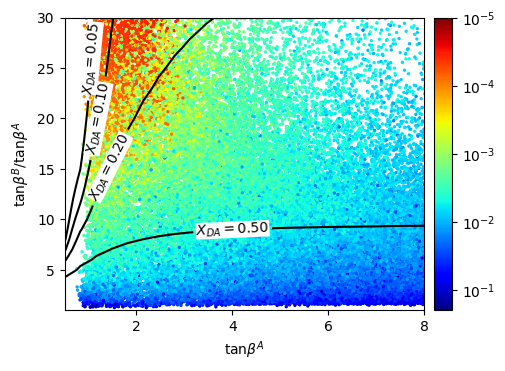}
    \label{fig:Fig2_350_II}
  \end{subfigure}
  \begin{subfigure}{0.495\textwidth}
    \centering
    \caption{Type Y, $v^B/v^A = 3.5$}
    \includegraphics[width=\textwidth]{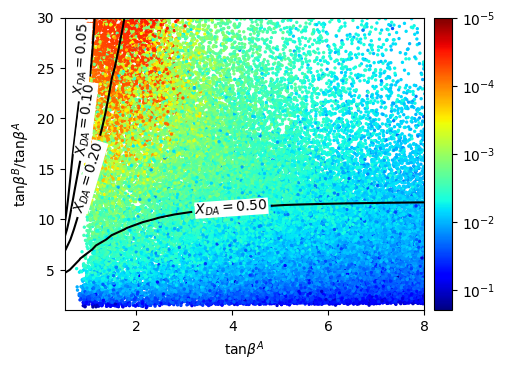}
    \label{fig:Fig2_350_Y}
  \end{subfigure}
  \captionsetup{justification=justified}
\caption{Contours of constant $X_{\text{DA}}$ for different $v^B/v^A$, different types, $r_T = 0.45$ and $m_{u^A} = 2.16$~MeV. The colour corresponds to the minimum of $T_{\theta_1}$ and $T_{\theta_2}$.}\label{fig:XDAcountours}
\end{figure}

\begin{figure}[t!]
  \centering
   \captionsetup{justification=centering}
    \begin{subfigure}{0.495\textwidth}
    \centering
    \caption{Type II, $v^B/v^A = 2.5$}
    \includegraphics[width=\textwidth]{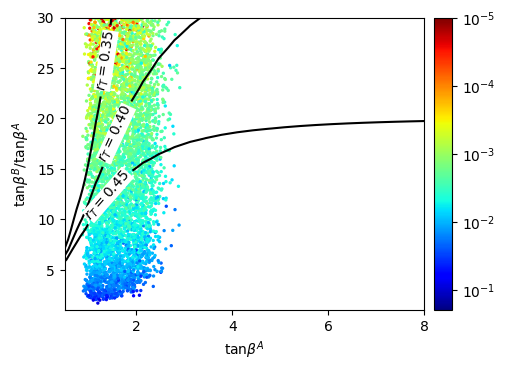}
    \label{fig:Fig3_250_II}
  \end{subfigure}
  \begin{subfigure}{0.495\textwidth}
    \centering
    \caption{Type Y, $v^B/v^A = 2.5$}
    \includegraphics[width=\textwidth]{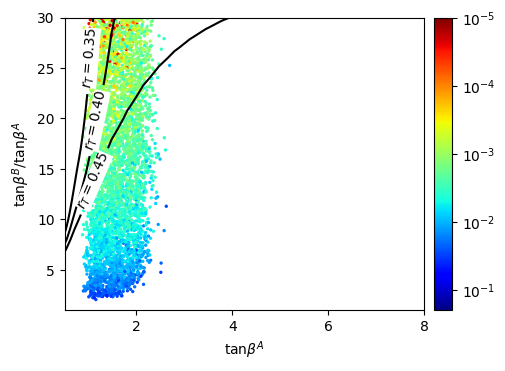}
    \label{fig:Fig3_250_Y}
  \end{subfigure}
  \vspace{0.0cm}
  \begin{subfigure}{0.495\textwidth}
    \centering
    \caption{Type II, $v^B/v^A = 3.0$}
    \includegraphics[width=\textwidth]{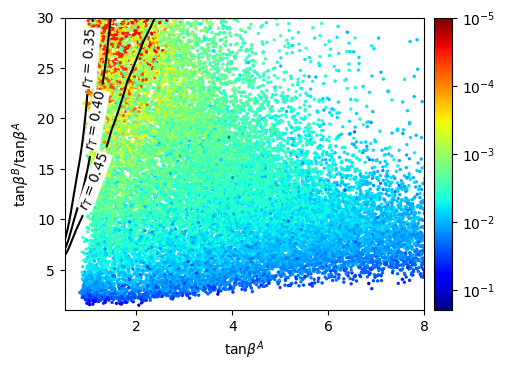}
    \label{fig:Fig3_300_II}
  \end{subfigure}
  \begin{subfigure}{0.495\textwidth}
    \centering
    \caption{Type Y, $v^B/v^A = 3.0$}
    \includegraphics[width=\textwidth]{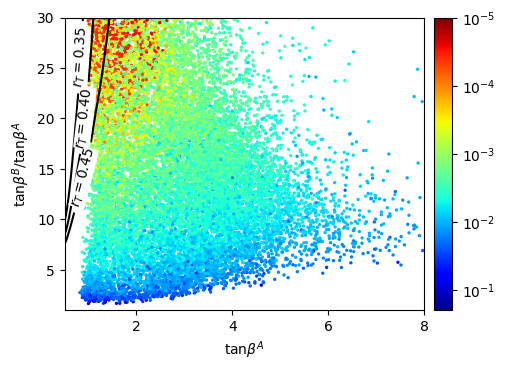}
    \label{fig:Fig3_300_Y}
  \end{subfigure}
  \vspace{0.0cm}
  \begin{subfigure}{0.495\textwidth}
    \centering
    \caption{Type II, $v^B/v^A = 3.5$}
    \includegraphics[width=\textwidth]{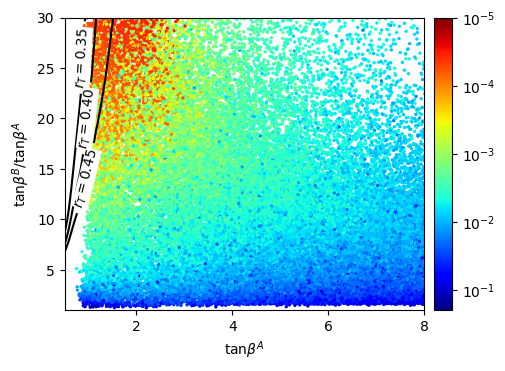}
    \label{fig:Fig3_350_II}
  \end{subfigure}
  \begin{subfigure}{0.495\textwidth}
    \centering
    \caption{Type Y, $v^B/v^A = 3.5$}
    \includegraphics[width=\textwidth]{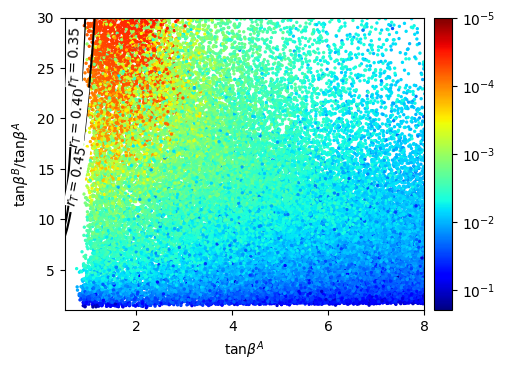}
    \label{fig:Fig3_350_Y}
  \end{subfigure}
  \captionsetup{justification=justified}
\caption{Contours of $X_{\text{DA}} = 10\%$ for different $r_T$, different $v^B/v^A$, different types and $m_{u^A} = 2.16$~MeV. The colour corresponds to the minimum of $T_{\theta_1}$ and $T_{\theta_2}$.}\label{fig:rTcountours}
\end{figure}

\begin{figure}[t!]
  \centering
   \captionsetup{justification=centering}
    \begin{subfigure}{0.495\textwidth}
    \centering
    \caption{Type II, $v^B/v^A = 2.5$}
    \includegraphics[width=\textwidth]{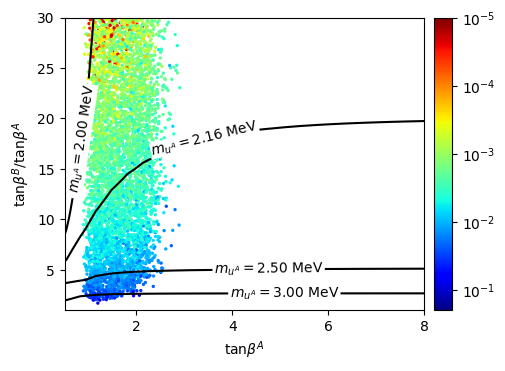}
    \label{fig:Fig4_250_II}
  \end{subfigure}
  \begin{subfigure}{0.495\textwidth}
    \centering
    \caption{Type Y, $v^B/v^A = 2.5$}
    \includegraphics[width=\textwidth]{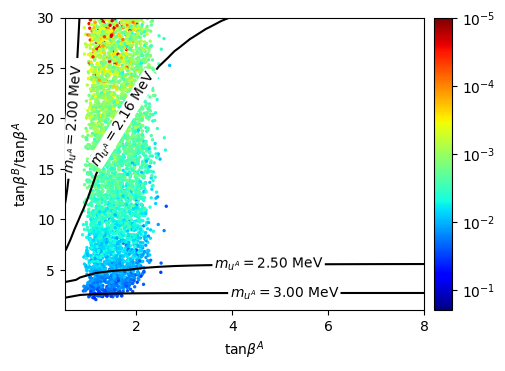}
    \label{fig:Fig4_250_Y}
  \end{subfigure}
  \vspace{0.0cm}
  \begin{subfigure}{0.495\textwidth}
    \centering
    \caption{Type II, $v^B/v^A = 3.0$}
    \includegraphics[width=\textwidth]{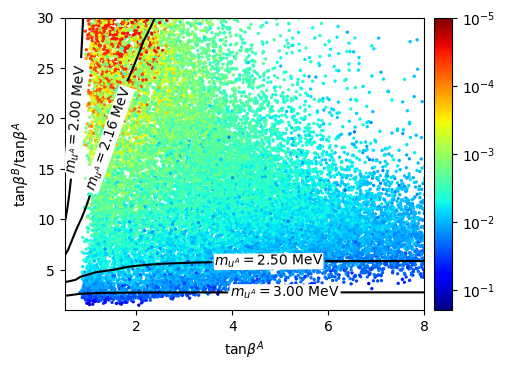}
    \label{fig:Fig4_300_II}
  \end{subfigure}
  \begin{subfigure}{0.495\textwidth}
    \centering
    \caption{Type Y, $v^B/v^A = 3.0$}
    \includegraphics[width=\textwidth]{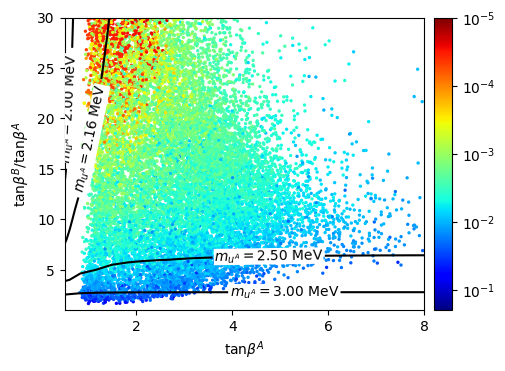}
    \label{fig:Fig4_300_Y}
  \end{subfigure}
  \vspace{0.0cm}
  \begin{subfigure}{0.495\textwidth}
    \centering
    \caption{Type II, $v^B/v^A = 3.5$}
    \includegraphics[width=\textwidth]{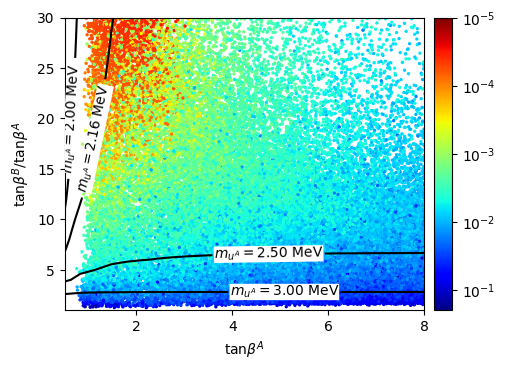}
    \label{fig:Fig4_350_II}
  \end{subfigure}
  \begin{subfigure}{0.495\textwidth}
    \centering
    \caption{Type Y, $v^B/v^A = 3.5$}
    \includegraphics[width=\textwidth]{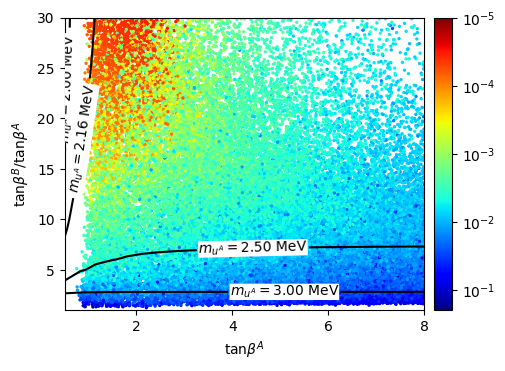}
    \label{fig:Fig4_350_Y}
  \end{subfigure}
  \captionsetup{justification=justified}
\caption{Contours of $X_{\text{DA}} = 10\%$ for different $m_{u^A}$, different $v^B/v^A$, different types and $r_T = 0.45$. The colour corresponds to the minimum of $T_{\theta_1}$ and $T_{\theta_2}$.}\label{fig:muAcountours}
\end{figure}

Having established all the necessary tools, it is now time to present the final results of this paper. They are shown in Figs.~\ref{fig:XDAcountours}, \ref{fig:rTcountours} and \ref{fig:muAcountours}. The colour of the points corresponds to the minimal value of $T_{\theta_1}$ and $T_{\theta_2}$.

First, Fig.~\ref{fig:XDAcountours} presents the contour of $X_{\text{DA}}$ for the central values of the masses of the up and down, an $r_T$ of 0.45 and three different values of $v^B/v^A$. Assuming the bound on $X_{\text{DA}}$ is really 10$\%$, these values indeed allow for mirror neutron dark matter. They however require either a very low $\tan\beta^A$ or a very large $\tan\beta^B/\tan\beta^A$. In addition, the tuning of the allowed points is at best of $\mathcal{O}(10^{-2})$, which is certainly worse than what one would hope for.

\begin{table}[t]
{\footnotesize
\setlength\tabcolsep{5pt}
\begin{center}
\scriptsize
\begin{tabular}{cccccccc}
\toprule
Figure                 & Contours               &  (a)  &   (b) &   (c) &   (d) &   (e) &   (f) \\
\cmrule
\ref{fig:XDAcountours} & $X_{\text{DA}} = 0.50$ &  ---  &  ---  &  ---  &  ---  &  ---  &  ---  \\
                       & $X_{\text{DA}} = 0.20$ &  ---  &  ---  &  ---  &  ---  &  ---  &  ---  \\
                       & $X_{\text{DA}} = 0.10$ &  ---  &  ---  &  ---  &  ---  &  5.36 &  1.91 \\
                       & $X_{\text{DA}} = 0.05$ &  ---  &  ---  & 10.14 &  1.90 &  1.88 &  1.33 \\
\cmrule
\ref{fig:rTcountours}  & $r_T = 0.45$           &  ---  &  ---  &  ---  &  ---  &  5.36 &  1.91 \\
                       & $r_T = 0.40$           &  ---  &  ---  &  5.43 &  1.85 &  1.91 &  1.35 \\
                       & $r_T = 0.35$           &  ---  &  2.00 &  1.73 &  1.26 &  1.30 &  1.04 \\
\cmrule
\ref{fig:muAcountours} & $m_{u^A} = 3.00$ MeV   &  ---  &  ---  &  ---  &  ---  &  ---  &  ---  \\
                       & $m_{u^A} = 2.50$ MeV   &  ---  &  ---  &  ---  &  ---  &  ---  &  ---  \\
                       & $m_{u^A} = 2.16$ MeV   &  ---  &  ---  &  ---  &  ---  &  5.36 &  1.91 \\
                       & $m_{u^A} = 2.00$ MeV   &  2.25 &  1.23 &  1.25 &  0.93 &  1.04 &  0.81 \\
\bottomrule
\end{tabular}
\end{center}
}
\caption{Maximum value of $\tan\beta^A$ than can accommodate $X_{\text{DA}}$ below 10$\%$ for each contour (or corresponding values in the case of Fig.~\ref{fig:XDAcountours}). An empty entry means that no such limit exists.} 
\label{tab:maxTanbetaA}
\end{table}

Second, Fig.~\ref{fig:rTcountours} presents coutours of $X_{\text{DA}} = 10\%$ for different values of $r_T$ and again using the central values of the masses of the up and down. As can be seen, the allowed region of parameter space shrinks as $r_T$ decreases. This is simply because the processes that convert mirror protons to mirror neutrons freeze earlier as $r_T$ decreases. A smaller $r_T$ also typically requires more tuning.

Finally, the results are crucially dependent on the masses of the light quarks, especially the up quark. As such, we show in Fig.~\ref{fig:muAcountours} contours of $X_{\text{DA}} = 10\%$ for different values of the mass of the up quark deviating from the central value by less than two sigmas. As can be seen, if the measurement of this mass were to fluctuate down considerably, only a very narrow band of $\tan\beta^A$ would be allowed in practice. Conversely, a small increase of this mass by even less than one sigma would make any value of $\tan\beta^A$ able to accommodate viable mirror neutron dark matter. It also allows far more points with small tuning.

One thing that might not be entirely clear from these plots is that, for a fixed $\tan\beta^A$, going to an arbitrarily large $\tan\beta^B/\tan\beta^A$ is not necessarily sufficient to bring $X_{\text{DA}}$ to an acceptable level. As such, the maximum $\tan\beta^A$ that can accommodate $X_{\text{DA}}$ below 10$\%$ (or corresponding values in the case of Fig.~\ref{fig:XDAcountours}) is presented in Table~\ref{tab:maxTanbetaA} for all relevant figures.

As a general rule, Type II performs better than Type Y. This is simply because, at large $\tan\beta^B/\tan\beta^A$, Type Y leads to a heavier mirror electron than Type II. This results in the processes that convert mirror protons to mirror neutrons freezing-out earlier.

\section{Conclusion}\label{Sec:Conclusion}
The Mirror Twin Higgs is a solution to the little hierarchy problem and also a natural setting for asymmetric dark matter. It however faces the problem that it leads to dark matter that mostly consists of mirror atoms, which is in severe conflict with bounds on dark matter self-interaction. This can be addressed by introducing a second Higgs doublet and its mirror copy, promoting the model to the Mirror Twin Two Higgs Doublet Model. The vevs can then be aligned to make the mirror proton heavier than the mirror neutron and in principle convert most mirror baryons to mirror neutrons, which would be cosmologically viable. Whether this mechanism is possible in practice is not trivial.

The goal of this paper was then to determine whether the MT2HDM can lead to viable mirror neutron dark matter and, if so, what amount of tuning is necessary. The end result is that it is indeed possible, but with some caveats.

The mirror BBN depends on the potential essentially only via the three parameters $v^B/v^A$, $\tan\beta^B/\tan\beta^A$ and $\tan\beta^A$. Decreasing $v^B/v^A$ decreases the amount of dark atoms, but, for sufficiently small values, Higgs signal strengths requirements force $\tan\beta^A$ to be very small and/or $\tan\beta^B/\tan\beta^A$ to be very large. Increasing the ratio $\tan\beta^B/\tan\beta^A$ decreases the amount of dark atoms, but requires additional tuning. Decreasing $\tan\beta^A$ also decreases the amount of dark atoms, but increases the amount of tuning required to obtain a large $\tan\beta^B/\tan\beta^A$.

In addition, increasing the precision of the Higgs signal strengths and the constraint on the effective number of relativistic degrees of freedom is bound to further constrain the model. In both cases, the regions of parameter space most likely to be probed next are those of low tuning, which, barring a deviation from SM predictions, would result in the model requiring more tuning. In practice, it would be difficult to exclude the limit of small $\tan\beta^A$ and large $\tan\beta^B/\tan\beta^A$, as unattractive as the tuning of this region might be.

Finally, the results are very dependent on the masses of the light quarks, which are unfortunately still poorly known. For the central measurement of the up quark mass, the mechanism is viable but requires tuning at the $\mathcal{O}(10^{-2})$ level. If this mass were to fluctuate down, the required tuning would drastically increase, making the model rather unattractive. In both of these cases, $\tan\beta^A$ would be confined to mostly small values, which could cause some complications for a UV completion. However, an upward fluctuation of even just one sigma would allow for tunings at the $\mathcal{O}(10\%)$ level, which is barely worse than what is required to pass the Higgs signal strengths requirements in the first place. In addition, it opens up the allowed range of $\tan\beta^A$, allowing for values well above 10. This additional freedom certainly facilitates UV completions and allows for values of $\tan\beta^A$ preferred by papers such as Ref.~\cite{Craig:2013fga, Badziak:2017syq, Badziak:2017kjk, Badziak:2017wxn}. As such, a more precise measurement of the masses of the light quarks could either effectively invalidate the model or, on the contrary, highlight it as a very natural explanation for dark matter.

\acknowledgments
HB would like to thank Yevgeny Kats for collaboration during the early stages of the project. HB is also grateful to Debjyoti Bardhan, Eric Kramer and Adi Zitrin for useful discussions. This research was supported in part by the Israel Science Foundation (grant no.\ 780/17) and the United States - Israel Binational Science Foundation (grant no.\  2018257).

\bibliography{biblio}
\bibliographystyle{utphys}

\end{document}